\documentclass[conference]{IEEEtran}
\IEEEoverridecommandlockouts
\usepackage{cite}
\usepackage{amsmath,amssymb,amsfonts}
\usepackage{algorithmic}
\usepackage{graphicx}
\newcommand{\argmax}{\mathop{\mathrm{argmax}}} 
\usepackage{hyperref}
\usepackage{multirow}
\usepackage{algorithm} 
\usepackage{algorithmic} 

\newlength\myindent
\newcommand\bindent{%
  \begingroup
  \setlength{\itemindent}{\myindent}
  \addtolength{\algorithmicindent}{\myindent}
}
\newcommand\eindent{\endgroup}

\usepackage{textcomp}
\usepackage{xcolor}
\def\BibTeX{{\rm B\kern-.05em{\sc i\kern-.025em b}\kern-.08em
    T\kern-.1667em\lower.7ex\hbox{E}\kern-.125emX}}
\begin{document}

\title{Deep Reinforcement Learning with Importance Weighted A3C for
QoE enhancement in Video Delivery Services\\

}

\author{\IEEEauthorblockN{Mandan Naresh, Paresh Saxena, Manik Gupta}
\IEEEauthorblockA{\textit{Computer Science \& Information Systems} \\
\textit{Birla Institute of Technology and Science Pilani}\\
Hyderabad, India \\
\{p20180420, psaxena, manik\}@hyderabad.bits-pilani.ac.in}
}

\maketitle

\begin{abstract}
Adaptive bitrate (ABR) algorithms are used to adapt the video bitrate based on the network conditions to improve the overall video quality of experience (QoE). Recently, reinforcement learning (RL) and asynchronous advantage actor-critic (A3C) methods have been used to generate adaptive bit rate algorithms and they have been shown to improve the overall QoE as compared to fixed rule ABR algorithms. However, a common issue in the A3C methods is the lag between behaviour policy and target policy. As a result, the behaviour and the target policies are no longer synchronized which results in suboptimal updates. In this work, we present ALISA: An Actor-Learner Architecture with Importance Sampling for efficient learning in ABR algorithms. ALISA incorporates importance sampling weights to give more weightage to relevant experience to address the lag issues with the existing A3C methods. We present the design and implementation of ALISA, and compare its performance to state-of-the-art video rate adaptation algorithms including vanilla A3C implemented in the Pensieve framework and other fixed-rule schedulers like BB, BOLA, and RB. Our results show that ALISA improves average QoE by up to 25\%-48\% higher average QoE than Pensieve, and even more when compared to fixed-rule schedulers.
\end{abstract}

\begin{IEEEkeywords}
Deep Reinforcement Learning, Video Delivery, Quality of Experience (QoE), Adaptive Bit Rates (ABR), Actor-critic methods

\end{IEEEkeywords}

\section{Introduction}
\label{intro}
There has been rapid growth in the usage of Internet-connected devices in recent years, and this trend is predicted to continue in the future. The authors of \cite{globalphenomenareport} report that video streaming accounted for 53.72\% of all internet traffic in the first half of 2021. Moreover, the number of IP network-connected devices are predicted to be thrice the global population by 2023 \cite{ciscoreport}, where HTTP-based video streaming will account for a large part of network traffic. However, several studies \cite{videodeliveryqualitystudy} have shown that low video quality often results in users abandoning video sessions, leading to considerable losses for content providers. AI has the ability to significantly improve a wide variety of mobile  services, including video streaming, online gaming, voice-over IP, smart home applications, and remote health monitoring. It can be used to optimize the quality of experience (QoE) of video streaming for users. Dynamic Adaptive Streaming over HTTP (DASH) \cite{dash} has established itself as a significant standard for streaming video content over the best-effort Internet. In general, adaptive bitrate (ABR) algorithms have been extensively investigated for their potential to improve the quality of experience in DASH-based video streaming \cite{abrsurvey}. ABR algorithms automatically adjust the video bitrate in response to network conditions such as buffer occupancy and observed throughput in order to give a greater quality of experience for the end users. However, these algorithms make decisions based on a predefined set of criteria and are frequently tailored for certain conditions. This makes it difficult to generalize such methods to the wide variety of network conditions that exist in today's ever-changing networks.

Reinforcement learning (RL) \cite{suttonbarto} is a subfield of machine learning concerned with how agents should take actions in an environment in order to maximize some notion of  cumulative reward. Several recent studies have investigated the integration of reinforcement learning approaches into video streaming \cite{pensieve, nancy} with a goal to achieve a high QoE. RL techniques with asynchronous advantage actor-critic (A3C) methods \cite{asyncrl} have demonstrated a number of advantages over ABR  algorithms based on fixed rules. Several  researchers \cite{pensieve, nancy} have used a vanilla A3C method to generate adaptive bit rates for the purpose of increasing the overall quality of experience. The A3C \cite{a3c} agent consists of multiple actors and a central learner with a critic. Each actor generates experience separately and concurrently based on its own behaviour policy. Individual experiences are then communicated to the central learner, which modifies the target policy (the policy that the A3C agent is attempting to learn) in response to the generated experience. However, A3C agents require a huge quantity of data to learn an appropriate policy. Increasing the number of actors is a common method for processing big amounts of data quickly. However, in such instances, each actor's behaviour policy begins to lag behind the target policy of the central learner \cite{impala2018}. As a result, the behaviour and target policies become out of sync, resulting in suboptimal updates. This may result into an inefficient use of bandwidth and a decrease in overall QoE while using RL for ABR algorithms. 

To address this issue, we integrate importance sampling weights \cite{impala2018} while using A3C methods for ABR generation to improve QoE for video streaming services. While assigning weights to the experience based on their relevance, our proposed approach solves the out-of-sync problem between behaviour and target policies and results in an overall higher QoE. Our solution is referred to as ALISA: Actor-Learner architecture with Importance Sampling for enhancing QoE in ABR algorithms. The proposed method is capable of generating adaptive bit rates via an actor-learner architecture based on reinforcement learning without relying on any pre-programmed model or assumption about the underlying systems. The current study makes a novel contribution by integrating importance sampling with A3C methods in order to train, learn, and generate adaptive bit rates while considering the distribution differences that may occur during training and, more importantly, when deploying the model in the real world. The main research contributions of this work are stated as follows:

\begin{itemize}
    \item Firstly, we present a new efficient ABR approach combining the importance sampling weights with actor-critic methods to improve video delivery services. By assigning importance sampling weights and, subsequently, allocating more significance to relevant experience, our method learns faster and gives an overall higher QoE than existing state-of-the-art ABR algorithms. Further, this helps the model to learn from samples with varying distributions in an efficient manner. 
    
    \item Second, we analyze the performance of the proposed approach using a widely-used Python-based framework and the MahiMahi simulator \cite{mahimahi}. We consider several datasets for performance evaluation utilizing traces from FCC \cite{fcc}, Norway \cite{norway}, OBOE \cite{OBOE}, and live video streaming \cite{LIVE}. We present a comprehensive study using three different variants of QoE metrics, $QoE_{lin}$, $QoE_{log}$, and $QoE_{HD}$, formulated as rewards for utilizing deep reinforcement learning. Finally, we also give a comparison over different network characteristics considering both lossless and lossy cases. 
    
   \item Third, we present the comparison of our proposed approach with other state-of-the-art ABR algorithms. This includes a comparison with the basic implementation of A3C, vanilla A3C (using the Pensieve framework) \cite{pensieve} and comparison with various non-RL ABR algorithms such as RB \cite{ratebased1}, BOLA \cite{bola}, RobustMPC \cite{mpc}, etc. Our results demonstrate that ALISA provides up to 25\%-48\% higher average QoE than vanilla A3C (Pensieve). However, the improvements are considerably bigger when compared to the fixed-rule schedulers.
    
\end{itemize}

The remaining paper is organized as follows. Section \ref{sec:relatedwork} presents the related work on ABR algorithms. Section \ref{section:background} presents the relevant background on reinforcement learning and actor-critic methods. Further, Section \ref{section:design} presents the problem statement, integration of importance sampling weights, proposed algorithm and system design. We present the experimental setup and results in Section \ref{section:measurementsetup} and Section \ref{sec:results}, respectively. Finally, we conclude our work in Section \ref{sec:conclusions}.

\section{Related Work}
\label{sec:relatedwork}



Several ABR algorithms have been developed \cite{buffbased,bola,mpc,pimentel2013qoe} to provide adaptive bit rates for video delivery over wireless networks. The algorithms can be characterized essentially as either rate-based or buffer-based. The rate-based algorithms \cite{ratebased1} predict the future chunk's bitrate as the maximum supported bitrate based on available network bandwidth and chunk history and the buffer-based algorithms predict based on the client's buffer occupancy \cite{buffbased,bola}. Due to the fact that the majority of these recommended techniques are based on pre-defined rules, they have a number of disadvantages. To begin, these algorithms are vulnerable to abrupt changes in network conditions, which might result in incorrect predictions. Second, while various approaches exist for achieving a higher QoE, each option has a trade-off. For instance, using the highest supported bitrate for each chunk may result in a loss of smoothness due to video resolution changes. Finally, the bitrate chosen for a current chunk frequently has an effect on the bitrate chosen for subsequent chunks. For instance, downloading chunks at the highest available bitrate may result in a reduction in the bit rate and quality of subsequent chunks in order to avoid rebuffering.

Recently, in addition to fixed-rules-based ABR algorithms, machine learning and deep learning have also been widely used to generate ABR algorithms. Model predictive control (MPC) \cite{mpc} has been used with deep learning in \cite{deepmpc} for more accurate throughput estimations. Further, a combination of machine learning, deep learning, and reinforcement learning is used in \cite{q2abr} to obtain improvements in QoE as compared to previous rule-based and ML-based approaches. The prediction of bitrate as a linear combination of input parameters is modeled in \cite{dlabr}. It uses a deep neural network to learn a suitable function. The deep learning model is also used in \cite{userexperienceabr} to learn the areas of interest in a video for a specific user to effectively allocate bitrate budgets. This methodology, along with the usual bandwidth and buffer occupancy, are jointly considered under the MPC framework to demonstrate an improvement over semantics-agnostic approaches.

Recently, there has been a focus on the development of a new class of ABR algorithms that make use of reinforcement learning. Numerous attempts have been made to apply Q-Learning for this task \cite{qlearning1,qlearning3}. However, these works employ a tabular Q-learning method, which makes expanding it to wider state spaces impossible. Additionally, the prediction of bitrate based solely on the most recently seen chunk is done in \cite{qlearning1}. It ignores the many most recently seen chunks that can enhance overall performance. To solve the issues of Q-learning in large state space, actor-critic methods for ABR generation are explored in \cite{pensieve,a3cabr1,nancy,deeprl,SAC-ABR}. In these papers, the A3C agent is used to generate ABRs and achieves a higher QoE than the majority of other fixed-rule-based ABR algorithms. However, the major issue with A3C is the lagging behind of an actor's behaviour policy as compared to the central learner's target policy \cite{impala2018,avga3c,a3cgs,a4c,ffe}. This has an effect on the performance of the A3C agent in existing reinforcement learning-based video distribution systems, resulting in decreased sample efficiency and the acquisition of a suboptimal policy. We propose and evaluate the integration of importance sampling weights to experiences depending on their relevance in order to solve a significant limitation of the existing A3C agent implementations for HTTP-based video delivery systems.

\section{Background}
\label{section:background}\label{background}

In this section, we present a brief overview of reinforcement learning and actor-critic methods.

\subsection{Reinforcement Learning}
A reinforcement learning solution \cite{suttonbarto} aims to learn a mapping from the state space to the action space by repeated interaction between the RL agent and the environment. The RL problem is modeled as a Markov decision process with states and actions. Let us consider a discrete system where at each time step $t \in \{0, 1, 2, ...\}$, the RL agent observes its state \(s_{t}\), takes an action \(a_{t}\), moves to state \(s_{t+1}\) and receives a reward $R(s_t, a_t)$ = \(r_{t+1}\). Further, for a sequence of states and actions, the discounted cumulative reward is defined as $R(\tau) = \sum_{k = 0}^{\infty} \gamma^{k}r_{t+k+1}$
where \(\tau\) is the sequence of states and actions, i.e. \(\{(s_{t}, a_{t}), (s_{t + 1}, a_{t + 1}), ...\}\) and $\gamma \leq 1$ is a discount factor. The agent selects action based on a policy, $\pi : \pi_{\theta}(s_{t}, a_{t}) \rightarrow [0,1]$,  
where $\pi_{\theta}(s_{t}, a_{t})$ is the
probability that action $a_{t}$ is taken in state $s_{t}$ and $\theta$ are the
policy parameters upon which the actions are based. Following the policy $\pi$, the value function $V(s)$ for a state $s$ is defined as $V(s) = \mathbb{E}_{\pi}[R(\tau) \vert S_{t} = s]$
. The goal of an RL agent is to find the optimal policy $\pi^{*}$ that maximizes the overall discounted reward. The optimal policy is given by, 
\begin{equation}
\pi^{*}(s_{t}) = \argmax_a[R(s_{t}, a_{t}) + \gamma V(s_{t+1})]
\end{equation}

Under this framework, we can have value-based methods which learns a value function mapping each state-action pair to a value. The action with the biggest value in a state becomes the optimal action to take. We can also have policy-based methods which directly optimize the policy function as explained in the next subsection.

\subsection{Actor-Critic Methods}
As an improvement to the value-based methods, \cite{a3c} presented actor-critic methods, which helps to speed up learning by reducing the variance of estimated quantities. An actor-critic method consists of two models. An actor that learns the optimal policy, and a critic, that approximates the value function (utility of a state-action pair). At each step $t$, we use the current state $s_t$ to predict the action $a_t$ by using the policy $\pi$. This also returns a reward $r_{t+1}$. Using this information, the critic now computes the value of the state-action pair, $\hat{q}_{w} (s_{t}, a_{t})$. The policy update with respect to its parameters \(\theta\) is defined in terms of the gradient operator $\nabla$ as follows,
\begin{equation}
\Delta \theta = \alpha \nabla_{\theta}\log \pi_{\theta}(s_{t}, a_{t})\hat{q}_{w}(s_{t}, a_{t})
\end{equation}
where \(\alpha\) is the actor learning rate, \(\hat{q}_{w}(s_{t}, a_{t})\) is the critic function that indicates how good an action \(a_{t}\) is in state \(s_{t}\). The parameters \(w\) of the critic function are updated as follows, 
\begin{equation}
\Delta w = \xi (R(s_{t}, a_{t}) + \gamma \hat{{q}}_{w}(s_{t+1}, a_{t+1})  - \hat{q}_{w}(s_{t}, a_{t})) \nabla_{w}\hat{q}_{w}(s_{t}, a_{t})
\end{equation}
where $\xi$ is the critic learning rate. 

However, a policy network trained in this manner may have high variance, which can cause instability during training. To mitigate this issue, the advantage actor-critic (A2C) \cite{a3c} framework introduces the advantage function to determine the advantage of the action taken in state \(s_{t}\) as compared to the average value of actions in \(s_{t}\). The advantage function is defined as the temporal difference (TD) error:
\begin{equation}
A(s_{t}, a_{t}) = R(s_{t}, a_{t}) + \gamma V(s_{t+1}) - V(s_{t})
\end{equation}
The final gradient-based update for the actor is as follows,
\begin{equation}
\Delta \theta = \alpha \nabla_{\theta}\log \pi_{\theta}(s_{t}, a_{t})A(s_{t}, a_{t}) + \beta \nabla_{\theta}H(\pi_{\theta}(.\vert s_{t}))
\end{equation}
where \(H(\pi_{\theta}(.\vert s_{t}))\) is the entropy factor which promotes random actions and \(\beta\) is the regularization term. The entropy term is defined as 
\begin{equation}
H(\pi_{\theta}(.\vert s_{t})) = -\sum_{a}\pi_{\theta}(a \vert s_{t})\log(\pi_{\theta}(a \vert s_{t}))
\end{equation}
The value of $\beta$ is initially set to a high value to promote exploration early on, and it is reduced as training progresses. To enhance training speed, Asynchronous Advantage Actor-Critic (A3C) framework \cite{a3c} is proposed to simulate multiple actors in parallel and asynchronously. These actors synchronize their parameters with the central learner at regular intervals. In our work, we use a modified A3C framework to generate ABR algorithms for  video delivery services.

\section{Problem Statement and Proposed Solution}
\label{section:design}

This section discusses the problem formulation and proposed solution for RL-based video distribution services with importance sampling, as well as the system design specifics.

\subsection{The Issue}

\begin{figure}[]
    \centering
    \includegraphics[width=0.4\textwidth]{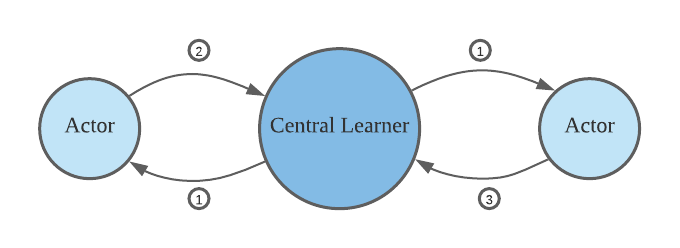}
    \caption{Illustration of A3C lagging issue. The figure shows three steps.  
        Step (1): Each actor synchronizes its weights with the central learner; Step (2): One of the actors provides experience to the central learner, which updates the weights of the target policy; 
        Step (3): The other actor provides experience to the central learner. The behaviour policy for the experience is not synchronized with the latest version of the target policy (updated in step 2), hence the experience is based on an older policy. 
        }
    \label{fig:a3cproblem}
\end{figure}

Reinforcement learning agents often require a large amount of experience to model the environment effectively and accurately. A common technique to achieve this goal is to increase the number of actors during the training. However, this strategy has an inherent flaw. As shown in Figure \ref{fig:a3cproblem}, the central learner first synchronizes its weights with those of all the actors (Step 1), after which the actors provide their experience to the central learner (Step 2). However, there are instances when some actors may delay sending the updates to the central learner. For example, as shown in Figure \ref{fig:a3cproblem}, the central learner updates its target policy even before receiving the experience from the actor on the right side (Step 3). Therefore, the behaviour policy associated with this event lags behind the target policy, and the experience may be less useful to update the current target policy. This issue is exacerbated further by the presence of more actors when actors start generating experience with an older version of the behaviour policy. The lack of synchronization between the behaviour and target policies results in suboptimal updates. Eventually, this leads to learning an overall suboptimal policy. We intend to develop approaches that compensate for the behaviour policy falling behind the target policy during training, allowing us to achieve higher performance on unseen test data.

\subsection{ALISA: System Design}

\begin{figure*}[]
    \centering
    \includegraphics[width=0.7\linewidth]{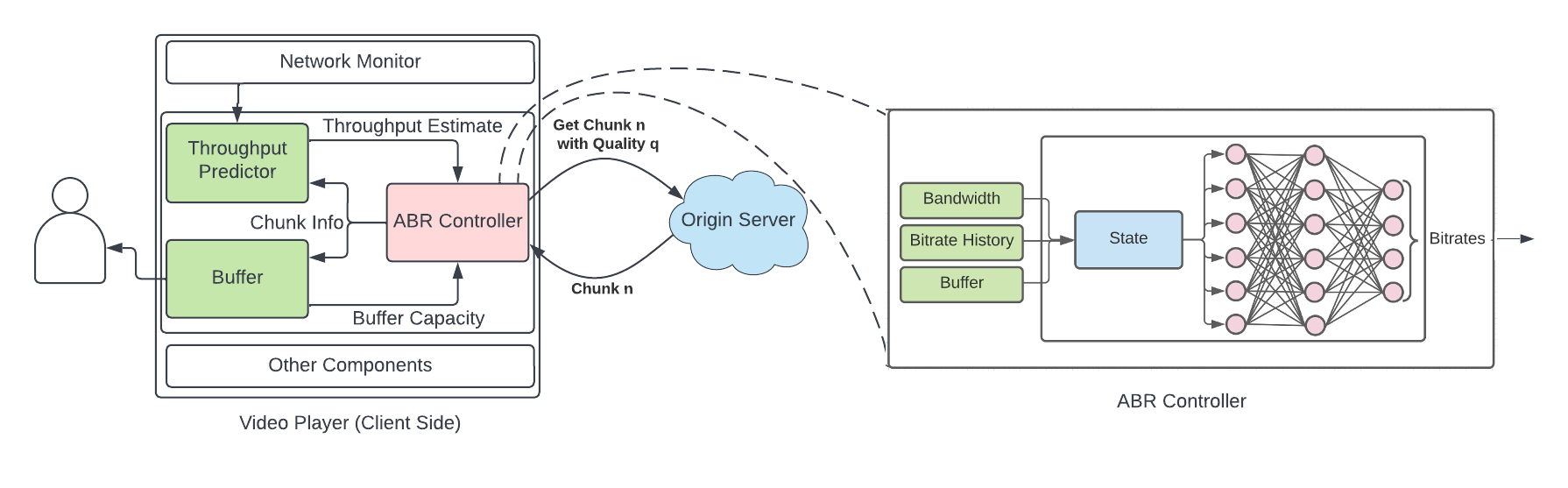}
    \caption{ALISA: System Design}
    \label{fig:alisaarch}
\end{figure*}

\begin{figure*}[]
    \centering
    \includegraphics[width=\linewidth]{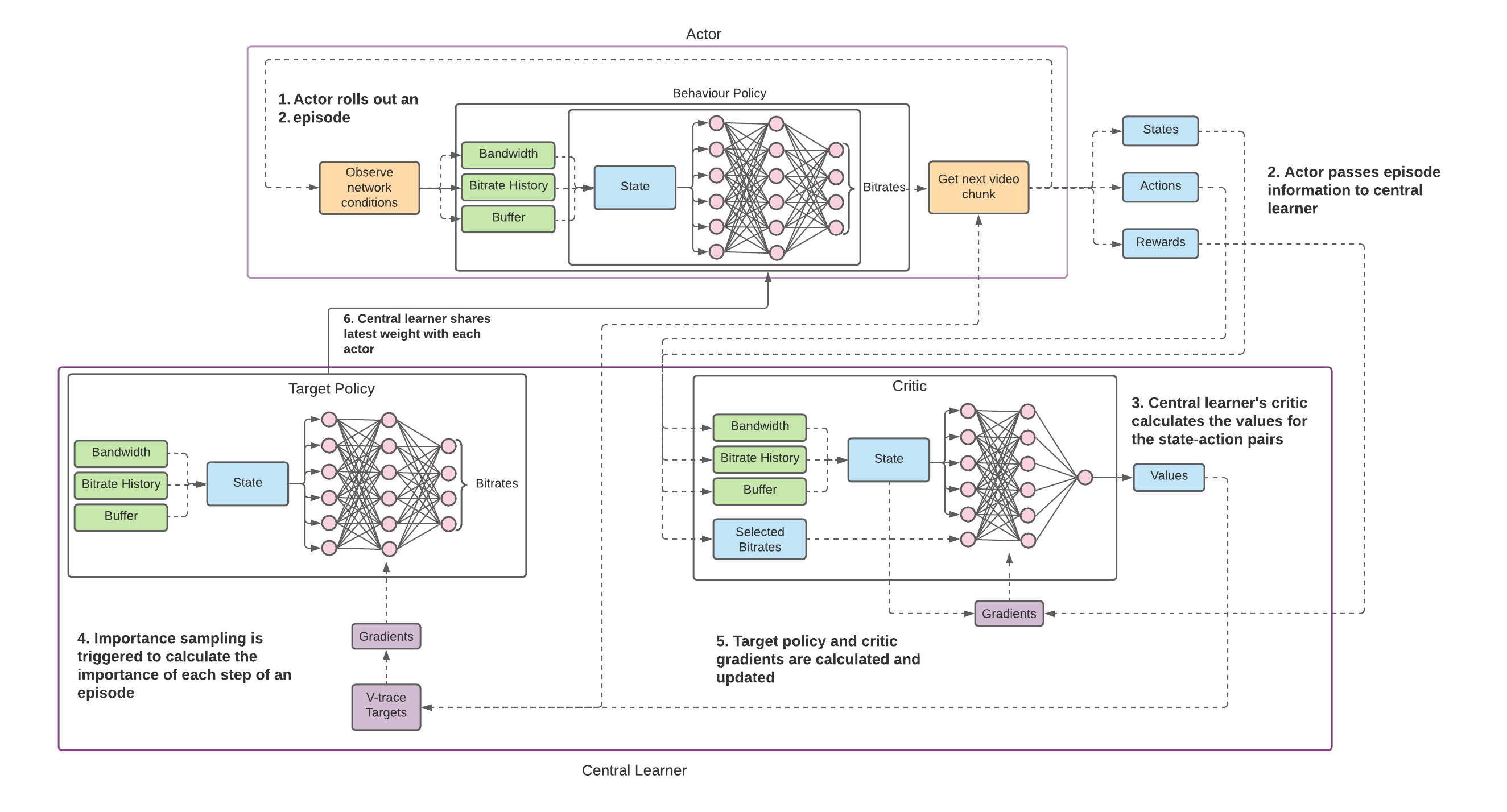}
    \caption{Detailed flow design for training the RL-based ABR controller of ALISA}
    \label{fig:alisatrain}
\end{figure*}

Our proposed solution ALISA builds on the DASH framework and uses deep RL-based A3C methods to achieve a higher QoE than existing state-of-the-art ABR algorithms for video streaming. Figure \ref{fig:alisaarch} presents the overall ALISA's system design. The user streams a video on their devices on a video player, whose main component is the ABR controller. It observes several state parameters on the client side, such as bandwidth, bitrate selection history, and buffer occupancy, and decides the action to take, i.e., the bitrate selection for the next chunk. At each step, it also observes some reward (QoE) as a result of its actions. 

We now describe the training process used by the ABR controller of ALISA. As shown in Figure \ref{fig:alisatrain}, the training environment is composed of multiple actors who are coordinated by a single central learner. The actor contains a behaviour policy as its parameters, while the central learner maintains the target policy and the critic parameters. The behaviour policy, the target policy, and the critic function are all modeled as neural networks. The training process can be considered as the repetition of the following steps until convergence:
\begin{enumerate}
\item First, an actor simulates an episode and generates a batch of experiences consisting of the states, the corresponding actions taken by the actor, and the rewards received as a result.
\item The experience is then passed back to the central learner.
\item The central learner calculates the values for each step of the experience using the critic parameters.
\item The central learner calculates the \(V-\)trace targets after incorporating the importance sampling strategy discussed in the next subsection. 
\item The critic gradients are computed using the observed states and their corresponding rewards, while the target policy gradients are computed using the observed states, the corresponding actions, the obtained rewards, and the \(V-\)trace targets. The target policy and the critic network are now updated using backpropagation.
\item Finally, the central learner shares the latest version of the target policy with each actor, which sets their behaviour policy to the newest target policy to generate the next batch of experiences.
\end{enumerate}

ALISA effectively decouples the acting and learning processes while also compensating for the resulting off-policy shift. This has significant implications for the development of ABR algorithms. Due to the vast volume of video being streamed to users worldwide, the ALISA architecture enables constant fine-tuning of the ABR algorithm and adaptation to ever-changing network conditions, all without jeopardising the users' privacy. While the streaming devices continuously make bitrate selections, the decisions can be relayed to the central learner located on a remote cloud server, where federated learning can take place \cite{federatedlearning}. The latest policy can be synchronized between the end-user devices and the remote server at regular intervals. This allows video streaming services to respond much faster to fluctuations in network conditions and changes in video streaming behaviour over time, allowing a new model to be retrained much faster. While we are limited by available data to demonstrate the benefit of ALISA on fine-tuning, we show in Section \ref{sec:results} how ALISA not only achieves a higher QoE but also does so in less than 50\% of the time required for comparable methods. We anticipate that this benefit observed in training the model from scratch will also extend while fine-tuning the model as we have suggested. 

\subsection{ALISA: Integration of Importance Sampling for Policy Update}
\label{section:importance sampling}
Importance sampling is a commonly used technique for resolving data distribution mismatches.  It provides the estimation of the expected value of a function \(f(x)\), where \(x\) follows a probability density function \(a\) on the domain \(\mathcal{D}\), by sampling values from a different distribution \(b\) on the same domain \(\mathcal{D}\) as,

\begin{equation}
\mathbb{E}(f(X)) = \mathbb{E}_{r}\left( \frac{f(X)a(X)}{b(X)}\right)
\label{eq:importance}
\end{equation}

Importance sampling alters the data collected from the distribution \(b\) in such a way that it looks to have been sampled from the distribution \(a\). This effectively addresses the distribution mismatch issue. In this work, we integrate importance sampling with ALISA to overcome the distribution mismatch between the target policy \(\pi\) and the behaviour policy \(\mu\). Correlating to the notation in Equation (\ref{eq:importance}), we have \(a \equiv \pi\) and \(b \equiv \mu\). Similar to the authors of \cite{impala2018}, we use the \(n\)-step \(V\)-trace target to adjust for the off-policy shift. The \(n\)-step \(V\)-trace target now acts as an estimate of the value function \(V\) for the target policy \(\pi\) using an older version of the behaviour policy \(\mu\). The $n$-step $V$-trace target is defined as,
\begin{equation}
v_{j} \doteq V(s_{j}) + \sum_{t=j}^{j+n-1}\gamma^{t-j}\left(\prod_{i=j}^{t-1}c_{i}\right)\delta_{t}V
\end{equation}
where,
\begin{itemize}

    \item \(\delta_{t}V \doteq \rho_{t}(r_{t}+\gamma V(s_{t+1}) - V(s_{t}))\) is the temporal difference.
    
    \item \(\rho_{t} \doteq min\left (\overline{\rho}, \frac{\pi(a_{t} \vert s_{t})}{\mu(a_{t} \vert s_{t})}  \right )\) and \(c_{i} \doteq min\left (\overline{c}, \frac{\pi(a_{i} \vert s_{i})}{\mu(a_{i} \vert s_{i})}  \right )\) are the importance sampling weights. The importance sampling weights \(\rho_{t}\) and \(c_{i}\) are used to give importance to experience, which is more relevant to the target policy than the behaviour policy. Here, $\pi$ denotes the target policy, and \(\mu\) denotes the behaviour policy.
    
    \item \(\overline{\rho}\) and \(\overline{c}\) are lower threshold values for their corresponding importance sampling weights, which we set to \(1\) throughout our work.

    \item  \(\overline{\rho_{t}}\) denotes how much more probable the action \(a_{t}\) taken in state \(x_{t}\) is according to the target policy compared to the behaviour policy.
    
    \item  \(\prod_{i=j}^{t-1}c_{i}\) denotes how much more probable the predicted path from state \(s_{j}\) to \(s_{t - 1}\) is according to the target policy compared to the behaviour policy.
\end{itemize}

 Subsequently, the \(V\)-trace targets are used in place of \(V\) for gradient computation. The $n$-step $V$-trace target can also be defined recursively as
\begin{equation}
v_{j} = V(s_{j}) + \delta_{j}V + \gamma c_{j}(v_{j+1} - V(s_{j+1}))
\end{equation}
 
 which we use during implementation throughout our work.
As a result of these calculations, actions that are more likely to be taken according to the target policy contribute more to the V-trace target. Hence the importance sampling weights help the reinforcement learning model to focus on the experience, which is more relevant and leads to better parameter updates and assigns less importance to suboptimal experience. Using the above definition of the V-trace target, Algorithm \ref{alg:1} outlines ALISA's policy update algorithm with the importance sampling weights where $n$ equals to the length of the episode. The $V$-trace target calculation with ALISA takes as input the information related to an episode consisting of the sequence of states ($s$), the sequence of action probabilities according to the behaviour policy ($a_{b}$), the sequence of rewards ($r$) along with the meta parameters $\overline{\rho}$ and $\overline{c}$ and the actor and critic models from Line $2$ to Line $8$. As a result, the $\rho$ is updated to a minimum of $\overline{\rho}$ and quotient of target policy and behaviour policy in Line $15$. In Line $16$ , the change to be added in the critic values for s ($V$) is computed . This is followed by the calculation of $c$ which is assigned as the minimum of $\overline{c}$ and $\rho$ in Line $17$. The \(V\)-trace targets are updated by adding the value computed in Line $16$ to the current \(V\)-trace targets. Line $19$ to Line $21$ explains how the \(V\)-trace targets for each \textbf{i} of the loop is updated.
As a consequence of importance sampling, actions that are more likely to be taken by the current target policy contribute more to the gradients compared to actions that are likely to be taken by earlier lagging versions of the target policy but not the current one. This algorithm guides the RL model to focus on the experience which matters more and assign less importance to other less relevant experiences.

\begin{algorithm}[]
\caption{Calculation of \(V\)-trace targets using ALISA for an episode}
\label{alg:1}
\begin{algorithmic}[1] 

\STATE \textbf{Input}:  
    \bindent
\STATE  {\textbf{\boldmath{$s$}}}: Sequence of states 
			\STATE 	\textbf{\boldmath{$a_{b}$}}: Sequence of action probabilities according to behaviour policy
              \STATE       \textbf{\boldmath{$r$}}: Sequence of rewards 
                   \STATE  \textbf{\boldmath{$\overline{\rho}$}}: Lower threshold for $\rho$
                   \STATE  \textbf{\boldmath{$\overline{c}$}}: Lower threshold for $c$ 
                   \STATE  \textbf{actor}: target policy from central learner
                 \STATE    \textbf{critic}: critic model
    \eindent
\STATE \textbf{Output}:
 \bindent
\STATE {$v$}: \(V\)-trace targets 
\STATE $V \leftarrow$ critic values for $s$ 
\STATE $p_{b} \leftarrow$ behaviour policy probabilities for optimal action 
\STATE $a_{t} \leftarrow$ target policy probabilities for $s$ 
\STATE $p_{t} \leftarrow$ target policy probabilities corresponding to optimal actions of behaviour policy, computed using $a_{t}$ and $p_{b}$ \\
\STATE $\rho \leftarrow min\left(\overline{\rho}, \frac{p_{t}}{p_{b}}\right)$ \\
\STATE $\delta_{t}V \leftarrow \rho \left( r + \gamma\ V_{+1} - V\right)$ \\
\STATE $c\leftarrow min\left(\overline{c}, \rho\right)$\\
\STATE $v \leftarrow V + \delta_{t}V$
\FOR{each position \textbf{i} from rear of v}
        \bindent
  \STATE       \(v_{i} \mathrel{+}= \gamma c_{i} * (v_{i + 1} - V_{i + 1})\)
        \eindent
\ENDFOR
\eindent

\end{algorithmic}
\end{algorithm}

\section{Experimental Details}\label{section:measurementsetup}

In this section, we describe the experimental setup together with the performance metrics used to evaluate ALISA's performance. 

\subsection{Experimental Setup}
\label{sec:experimentalsetup}
We use the Python-based framework proposed in \cite{pensieve}, to generate and test our ABR algorithms. The client requests chunks of data from the server and provides it with parameters pertaining to the observed network conditions like bandwidth, buffer occupancy, and bitrate history. We have integrated importance sampling as described in Section 
\ref{section:design} and assigned weights to the target and the behaviour policies. To emulate network conditions for effectively testing our trained RL model, the MahiMahi \cite{mahimahi} framework has been used, which is a record-and-replay HTTP framework for this task. 

We use four datasets as part of our training set: the broadband dataset provided by the FCC \cite{fcc}, the mobile dataset collected in Norway \cite{norway}, the OBOE traces \cite{OBOE}, and the live video streaming dataset \cite{LIVE}, which have been pre-processed according to the MahiMahi format. Subsequently, we compare ALISA with the following state-of-the-art ABR algorithms:
\begin{itemize}
\item Pensieve (Vanilla A3C) \cite{pensieve}: uses vanilla A3C without any additional techniques to train the agent for delivering adaptive bit rates.  
\item Rate-Based (RB) \cite{ratebased1}: RB predicts the maximum supported bitrate based on the harmonic mean of past observed throughput.
\item Buffer-based (BB) \cite{buffbased}: BB selects bitrate based on client's buffer occupancy.
\item BOLA \cite{bola}: Bitrate selection is made exclusively based on buffer occupancy using Lyapunov optimization. 
\item RobustMPC \cite{mpc}: MPC uses buffer occupancy observations and throughput predictions similar to RB. Additionally, RobustMPC accounts for errors between predicted and observed throughputs by normalizing throughput estimates.
\end{itemize}
To quantify the performance of the ABR algorithms, we use the formulation of QoE:
\begin{equation}
QoE = \sum_{n=1}^{N}q(b_{n}) - \mu \sum_{n=1}^{N}T_{n} - \sum_{n=1}^{N-1} \left\vert q(b_{n+1})-q(b_{n})\right\vert
\label{eq:qoe}
\end{equation}
where $b_{i}$ and $q(b_{i})$ represent the bit-rate and quality, respectively, for chunk $i$. A higher bit rate means higher quality and a higher QoE. However, there are also penalties due to rebuffering time $T_{i}$ (represented by the second term) and fluctuations in video quality (represented by the final term) that hinders the overall smoothness. In this paper, we evaluate the performance of the proposed approaches with three QoE variants \cite{pensieve} that depend on the above general QoE metric: 
\begin{itemize}
    \item \(QoE_{lin}\): \(q(b_{n}) = b_{n}\) where value of rebuffer penalty is \(\mu = 4.3\),
    
    \item \(QoE_{log}\): \(q(b_{n})=log(b/b_{min})\) that considers the fact that the marginal improvement in quality decreases at higher bitrates with \(\mu = 2.66\) and
    
    \item \(QoE_{HD}\): assigns a higher value to higher quality bitrates and lower values to lower quality bitrates with $\mu=8$.
\end{itemize}

\subsection{Dataset details}
We use the following data sets for training, validation, and testing. Our selection of data sets for training, validation, and testing is in line with the previous experimental setups in \cite{pensieve}, \cite{OBOE}, \cite{LIVE}. For training, we have used three different data sets. The first data set consists of 127 traces, out of which 59 belongs to the FCC \cite{fcc} dataset while the remaining 68 belong to the Norway HSDPA  \cite{norway} dataset. The second data set consists of 428 OBOE traces \cite{OBOE} and the third data set consists of 100 live video streaming traces \cite{LIVE}. To demonstrate the benefits of ALISA over different trained models, we have generated three different trained model corresponding to three different data sets described above. For all three trained models, we have used the same validation data set, i.e., 142 Norway traces. Finally, for all three trained models, after validation, the testing is performed using 205 traces from the FCC dataset and 250 traces from the Norway HSDPA dataset. 

\begin{figure}
    \centering
    \includegraphics[width=0.45\textwidth]{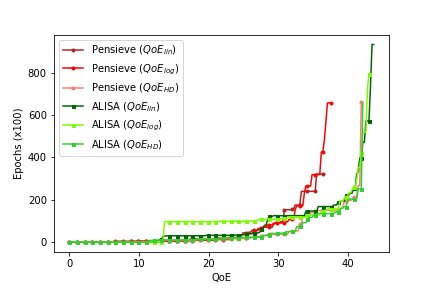}
    \caption{Maximum training QoE obtained versus epochs elapsed. ALISA is able to obtain a higher QoE faster for all three variants of the QoE metric.}
    \label{fig:convergence}
\end{figure}

\subsection{Training Methodology}
We train the three models for each configuration on Pensieve and ALISA, one each for \(QoE_{lin}\), \(QoE_{log}\) and \(QoE_{HD}\) as reward metrics. For each model, we use a consistent set of hyperparameters throughout. The discount factor \(\gamma\) is set to \(0.99\). The learning rates are set to \(0.0001\) and \(0.001\) for the actor and critic, respectively. In Pensieve,  the entropy
factor ($H(.)$) is controlled by using an entropy regularization factor  $(\beta)$. The $\beta$ uses entropy decay values from 1 to 0.1 over 100,000 epochs. In ALISA, we set both importance sampling thresholds \(\bar{\rho}\) and \(\bar{c}\) to 1. We train multiple models for different configurations of entropy weights. First, we train several models with a constant entropy weight for 100,000 epochs. Next, we use a decaying entropy weight where the entropy is gradually decreased over 100,000 epochs.
\subsection{Testing Methodology}
We select the model with the highest validation QoE for testing. We perform testing under both lossless and lossy conditions simulated using the MahiMahi \cite{mahimahi} framework. We perform tests under packet loss percentages of 0\%, 0.1\%, 0.5\%, 1\%, and 2\%, where random packets are dropped from the video stream. We evaluate all models on the three different QoE metrics discussed in Section \ref{sec:experimentalsetup}.

\section{Results}\label{sec:results}
In this section, we present the results and comparison of ALISA with other state-of-the-art ABR algorithms. 

\subsection{Convergence Speed}
ALISA takes advantage of the importance sampling strategy during training. As a result, it is often able to achieve a higher QoE compared to Pensieve (Vanilla A3C) in a shorter time. Figure \ref{fig:convergence} presents the plots of the epochs elapsed versus the maximum QoE achieved till then. These plots are generated during the training using the first data set, i.e., 127 traces from FCC and Norway data sets. Our results show that by the time ALISA obtains a high QoE of over 40, Pensieve is only able to obtain the highest QoE of approximately 35. Furthermore, ALISA achieves a QoE over 40 in less than $1/3^{rd}$ of the time required for Pensieve to achieve its highest QoE. This demonstrates ALISA's advantage in learning and adapting to newer conditions faster, resulting in shorter training times. Similar results are observed during the training using OBOE and live video streaming traces.

\begin{table*}[]
\centering
\caption{Average QoE after training ALISA with all the three datasets and all the three variants of QoE metrics.}

\begin{tabular}{|l|c|c|c|c|c|c|c|c|c|}
\hline
\multicolumn{1}{|c|}{\multirow{2}{*}{Entropy values}} & \multicolumn{3}{c|}{FCC and Norway Traces}            & \multicolumn{3}{c|}{OBOE Traces}                 & \multicolumn{3}{c|}{Live Video Streaming Traces} \\ 
\multicolumn{1}{|c|}{}                                & \(QoE_{lin} \)  & \(QoE_{log}\)  &\( QoE_{HD}\)    &\( QoE_{lin}\)   & \(QoE_{log}\)   & \(QoE_{HD}\)    & \(QoE_{lin} \)  & \(QoE_{log} \)  & \(QoE_{HD}\)    \\ \hline
0 ($\times5$)                                                & 13.61          & 13.62         & 13.57          & 30.76          & 30.8           & 39.51          & 13.61          & 30.75          & 29.97          \\ \hline
0.001 ($\times5$)                                            & 13.61          & 13.62         & 13.61          & 42.47          & 30.78          & 39.52          & 13.61          & 30.87          & 13.57          \\ \hline
0.01 ($\times5$)                                             & 13.61          & 13.61         & 13.61          & 38.79          & 39.25          & 39.44          & 30.53          & 30.93          & 37.84          \\ \hline
0.1, 2, 0.1, 2, 0.1                                   & 41.57          & 38.91         & 42.21          & 41.43          & 44.03          & 38.4           & 43.27          & 38.54          & 41.36          \\ \hline
0.1 ($\times5$)                                              & 43.88          & 43.29         & 37.88          & 42.89          & 38.37          & 38.55          & 44.1           & 30.76          & 27.96          \\ \hline
0.25 ($\times5$)                                             & 40.91          & 43.29         & 37.61          & 42.85          & 43.2           & 39.57          & 43.22          & 44.03          & 37.69          \\ \hline
0.5 ($\times5$)                                              & 38.11          & 37.14         & \textbf{42.42} & 36.79          & 37.99          & 39.17          & 37.86          & 38.35          & 41.71          \\ \hline
0.75 ($\times5$)                                             & 31.11          & 32.11         & 41.99          & 29.85          & 30.69          & 40.97          & 32.23          & 32.83          & 42.28          \\ \hline
1, 0.75, 0.5, 0.25, 0.1                               & 43.22          & 44.09         & 42.4           & \textbf{44.34} & \textbf{44.94} & 40.61          & \textbf{44.76} & \textbf{45.96} & 41.66          \\ \hline
1 ($\times5$)                                                & 27.46          & 25.8          & 40.98          & 25.07          & 25.15          & 41.11          & 26.41          & 26.07          & 41.17          \\ \hline
2, 1.5, 1, 0.5, 0.1                                   & 43.86          & 44.19         & 41.27          & 43.16          & 43.73          & \textbf{41.16} & 44.45          & 45.54          & 41.76          \\ \hline
3, 2, 1, 0.5, 0.1                                     & \textbf{43.92} & 44.31         & 42.11          & 43.95          & 43.36          & 40.56          & 44.25          & 44.54          & 42.32          \\ \hline
4, 2, 1, 0.5, 0.1                                     & 43.15          & 43.4          & 41.8           & 42.99          & 44.93          & 39.33          & 43.92          & 45.93          & \textbf{42.66} \\ \hline
5, 2, 1, 0.5, 0.1                                     & 41.94          & \textbf{44.8} & 42.06          & 43.23          & 42.33          & 40.99          & 43.52          & 45.12          & 41.64          \\ \hline
\end{tabular}
\label{tab:trainingresults}

\end{table*}

\begin{table}[]
\centering
\caption{Average QoE achieved on all the datasets with all three variants of QoE metrics under emulation with no packet losses.}
\resizebox{8cm}{!}{
\begin{tabular}{|c|c|c|c|c|c|c|c|c|c|}
\hline
\multirow{2}{*}{Algorithm} & \multicolumn{3}{c|}{FCC and Norway} & \multicolumn{3}{c|}{OBOE} & \multicolumn{3}{c|}{Live video streaming} \\ 
                           & Linear      & Log       & HD         & Linear    & Log      & HD        & Linear        & Log          & HD           \\ \hline
ALISA                      & \textbf{43.03}       & \textbf{42.37}     & \textbf{256.29}     & \textbf{42.5}      & \textbf{41.79}    & \textbf{237.27}    & \textbf{46.57}         & \textbf{44.36}        & 228.63       \\ \hline
Pensieve                   & 39.62       & 35.26     & 239.08     & 37.52     & 37.01    & 194.29    & 39.12         & 41.68        & \textbf{234.72}       \\ \hline
BB                         & 12.02       & 12.78     & 84.24      & 14.08     & 20.00       & 80.36     & 13.81         & 20.26        & 63.08        \\ \hline
RB                         & 35.62       & 36.45     & 139.82     & 36.15     & 37.97    & 138.02    & 37.44         & 37.35        & 120.52       \\ \hline
BOLA                       & 34.25       & 35.3      & 141.04     & 35.04     & 37.09    & 139.1     & 35.82         & 36.05        & 121.02       \\ \hline
RobustMPC                  & 39.93       & 40.44     & 195.52     & 40.21     & 38.03    & 188.65    & 40.59         & 38.99        & 177.58       \\ \hline
\end{tabular}}

\label{tab:testresultslossless}
\end{table}

\begin{table}[hbt!]
\centering
\caption{Average QoE achieved on all the datasets with all the three variants of QoE metrics under emulation of random packet drops with 0.1\% probability.}
\resizebox{8cm}{!}{
\begin{tabular}{|c|c|c|c|c|c|c|c|c|c|}
\hline
\multirow{2}{*}{Algorithm} & \multicolumn{3}{c|}{FCC and Norway} & \multicolumn{3}{c|}{OBOE} & \multicolumn{3}{c|}{Live video streaming} \\  
                           & Linear      & Log       & HD         & Linear    & Log      & HD        & Linear        & Log          & HD           \\ \hline
ALISA                      & \textbf{44.58}       & \textbf{43.75}     & \textbf{244.38}     & \textbf{44.46}     & \textbf{42.07}    & \textbf{237.91}    & \textbf{45.76}         & \textbf{41.89}        & 233.96       \\ \hline
Pensieve                   & 41.99       & 35.06     & 235.12     & 39.34     & 36.69    & 194.37    & 36.62         & 39.08        & \textbf{241.10}       \\ \hline
BB                         & 16.62       & 20.97     & 73.16      & 17.08     & 19.81    & 0.64      & 15.40         & 19.58        & 68.76        \\ \hline
RB                         & 38.41       & 38.73     & 132.98     & 39.05     & 37.64    & 35.58     & 37.15         & 37.65        & 130.14       \\ \hline
BOLA                       & 37.58       & 37.21     & 129.27     & 37.84     & 36.03    & 34.81     & 35.95         & 36.08        & 126.24       \\ \hline
RobustMPC                  & 41.90       & 37.30     & 189.53     & 42.80     & 37.97    & 187.43    & 40.97         & 37.56        & 182.75       \\ \hline
\end{tabular}}

\label{tab:testresults.1}
\end{table}

\begin{table}[hbt!]
\centering
\caption{Average QoE achieved on all the datasets with all the three variants of QoE metrics under emulation with 0.5\% probability.}
\resizebox{8cm}{!}{
\begin{tabular}{|c|c|c|c|c|c|c|c|c|c|}
\hline
\multirow{2}{*}{Algorithm} & \multicolumn{3}{c|}{FCC and Norway Traces} & \multicolumn{3}{c|}{OBOE Traces} & \multicolumn{3}{c|}{Live video streaming Traces} \\  
                           & Linear      & Log       & HD         & Linear    & Log      & HD        & Linear        & Log          & HD           \\ \hline
ALISA                      & 37.34       & \textbf{43.86}     & \textbf{236.15}     & 39.2      & \textbf{42.24}    & \textbf{231.37}    & \textbf{42.53}         & \textbf{43.33}        & 227.66       \\ \hline
Pensieve                   & 35.43       & 34.1      & 214,59     & 36.44     & 36.22    & 201.32    & 34.98         & 40.80         & \textbf{231.19}       \\ \hline
BB                         & 10.59       & 18.74     & 66.13      & 13.97     & 18.79    & 69.6      & 12.12         & 18.41        & 64.51        \\ \hline
RB                         & 32.91       & 35.89     & 108.86     & 34.19     & 35.35    & 117.44    & 33.99         & 34.92        & 111.23       \\ \hline
BOLA                       & 32.39       & 34.59     & 108.98     & 34.12     & 33.97    & 112.67    & 33.78         & 33.77        & 110.80        \\ \hline
RobustMPC                  & \textbf{37.99}       & 38.2      & 177.87     & \textbf{39.48}     & 38.46    & 184.96    & 38.62         & 38.09        & 179.55       \\ \hline
\end{tabular}}

\label{tab:testresults.5}
\end{table}

\begin{table}[hbt!]
\centering
\caption{Average QoE achieved on all the datasets with all the three variants of QoE metrics under emulation with 1\% probability.}
\resizebox{8cm}{!}{
\begin{tabular}{|c|c|c|c|c|c|c|c|c|c|}
\hline
\multirow{2}{*}{Algorithm} & \multicolumn{3}{c|}{FCC and Norway Traces} & \multicolumn{3}{c|}{OBOE Traces} & \multicolumn{3}{c|}{Live video streaming Traces} \\ 
                           & Linear      & Log       & HD         & Linear    & Log      & HD        & Linear        & Log          & HD           \\ \hline
ALISA                      & \textbf{34.87}       & \textbf{38.18}     & \textbf{198.34}     & \textbf{35.35}     & \textbf{44.86}    & \textbf{181.71}    & \textbf{36.97}         & \textbf{38.2}         & 187.68       \\ \hline
Pensieve                   & 29.68       & 29.22     & 174.53     & 29.04     & 30.18    & 138.71    & 29.50         & 29.22        & \textbf{188.78}       \\ \hline
BB                         & 2.90        & 10.11     & 19.07      & 2.91      & 10.57    & 12.67     & 6.43          & 10.11        & -15.75       \\ \hline
RB                         & 27.62       & 28.58     & 78.17      & 27.38     & 28.72    & 77.99     & 28.00         & 28.58        & 25.38        \\ \hline
BOLA                       & 26.93       & 27.74     & 77.30      & 27.59     & 27.95    & 77.12     & 26.89         & 27.74        & 25.00        \\ \hline
RobustMPC                  & 32.92       & 33.13     & 149.21     & 32.62     & 33.47    & 142.51    & 33.74         & 33.13        & 145.87       \\ \hline
\end{tabular}}

\label{tab:testresults1}
\end{table}

\begin{table}[h!] 
\centering
\caption{Average QoE achieved on all the datasets with all the three variants of QoE metrics under emulation with 2\% probability.}
\resizebox{8cm}{!}{
\begin{tabular}{|c|c|c|c|c|c|c|c|c|c|}
\hline
\multirow{2}{*}{Algorithm} & \multicolumn{3}{c|}{FCC and Norway Traces} & \multicolumn{3}{c|}{OBOE Traces} & \multicolumn{3}{c|}{Live video streaming Traces} \\ 
                           & Linear      & Log       & HD         & Linear    & Log      & HD        & Linear        & Log          & HD           \\ \hline
ALISA                      & \textbf{27.60}       & \textbf{28.98}     & \textbf{128.53}     & \textbf{25.62}     & \textbf{26.67}    & \textbf{122.05}    & \textbf{28.73}         & \textbf{30.83}        & \textbf{122.47}       \\ \hline
Pensieve                   & 24.28       & 21.48     & 111.57     & 22.35     & 22.98    & 82.05     & 22.45         & 28.56        & 121.41       \\ \hline
BB                         & -13.44      & -6.74     & -41.18     & -15.48    & -6.28    & -37.73    & -14.49        & -4.15        & -38.48       \\ \hline
RB                         & 16.09       & 17.35     & 13.44      & 17.27     & 17.75    & 14.88     & 17.21         & 18.22        & 14.21        \\ \hline
BOLA                       & 17.89       & 18.59     & 16.05      & 18.08     & 18.01    & 16.08     & 17.48         & 18.76        & 16.49        \\ \hline
RobustMPC                  & 24.43       & 20.72     & 99.98      & 25.28     & 21.66    & 99.14     & 24.98         & 21.69        & 97.09        \\ \hline
\end{tabular}}

\label{tab:testresults2}
\end{table}

\subsection{Comparison with state-of-the-art ABR algorithms}
\subsubsection{Results with training and validation data sets}
We perform a comprehensive set of training on the three data sets and report our results on $QoE_{lin}$, $QoE_{log}$, and $QoE_{HD}$ metrics. Table \ref{tab:trainingresults} presents the rewards obtained after training and validation on the FCC and Norway traces OBOE traces and the live video streaming traces. We also investigate the use of different values for entropy regularization  $\beta$. Specifically, we consider the following values: 0, 0.001, 0.01, 0.1, 0.25, 0.5, 0.75, and 1. We have also explored the use of variations in $\beta$ during the training. For example, in Table \ref{tab:trainingresults}, $\{2,1.5,1,0.5,0.1\}$ refers to the scenario where $\beta = 2$ for the first 20000 epochs, $\beta = 1.5$ for the next 20000 epochs, and so on, till $\beta=0.1$ for the last 20000 epochs. Similarly, we have also used constant entropy regularization, where $0.1 \times 5$, in Table \ref{tab:trainingresults}, refers to the scenario where $\beta=0.1$ for all 100000 iterations. 

We note that the models do not converge well on training with very low or very high entropy. We found a constant entropy of 0.1 to work well for $QoE_{lin}$ and $QoE_{log}$ metrics, while 0.75 worked well for $QoE_{HD}$. We have also trained multiple times using a decaying entropy regularization. We start from a high value and gradually decrease our entropy weight every 20,000 epochs, gradually going down to 0.1. We find that decaying entropy regularization is more effective in almost all the cases, as seen from Table \ref{tab:trainingresults} since after a few epochs, a high exploration is not required to achieve an optimal policy. 

\begin{figure}
\centering
\includegraphics[width=0.5\textwidth]{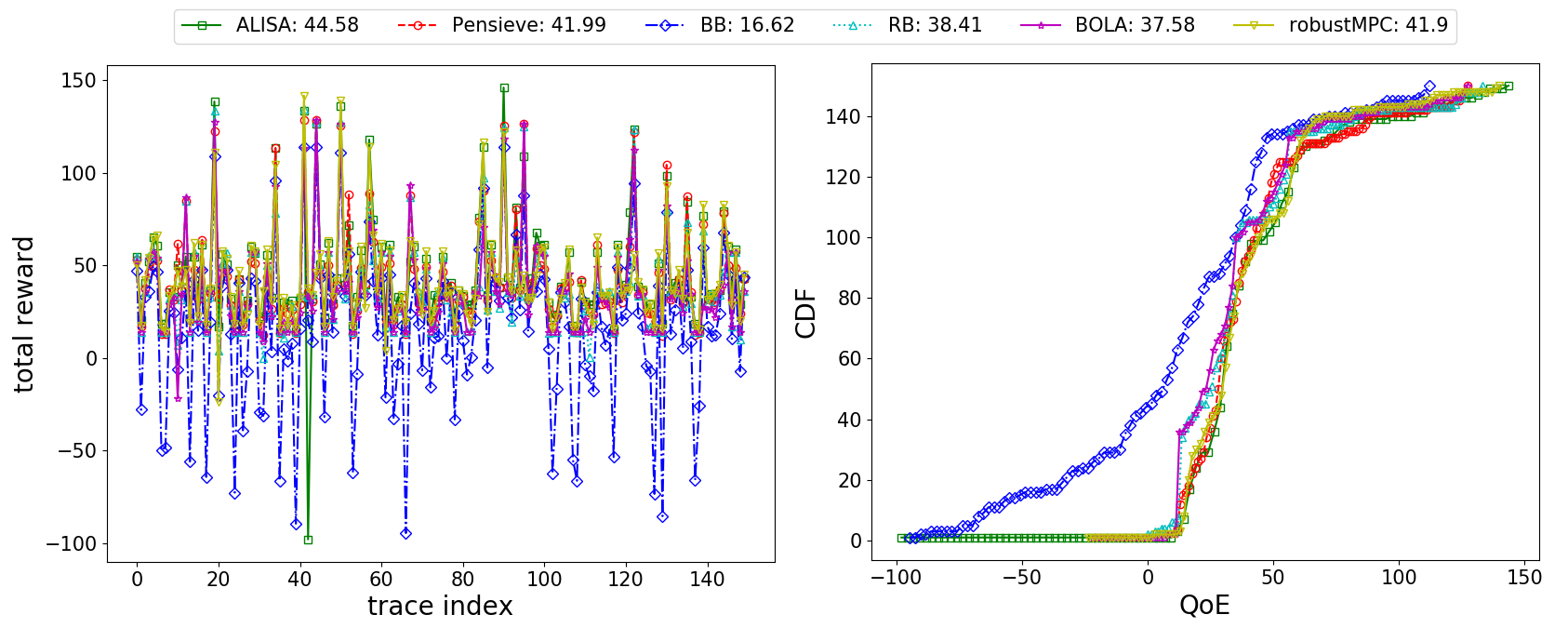}
\caption{Comparison of ALISA over other ABR algorithms with the \(QoE_{linear}\) metric: (left) average reward overall test traces; (right) CDF vs QoE plot  under emulation of random packet drops with 0.1\% probability.}
\label{fig:rewardcdf}
\end{figure}

\begin{figure}
\centering
\includegraphics[width=0.35\textwidth]{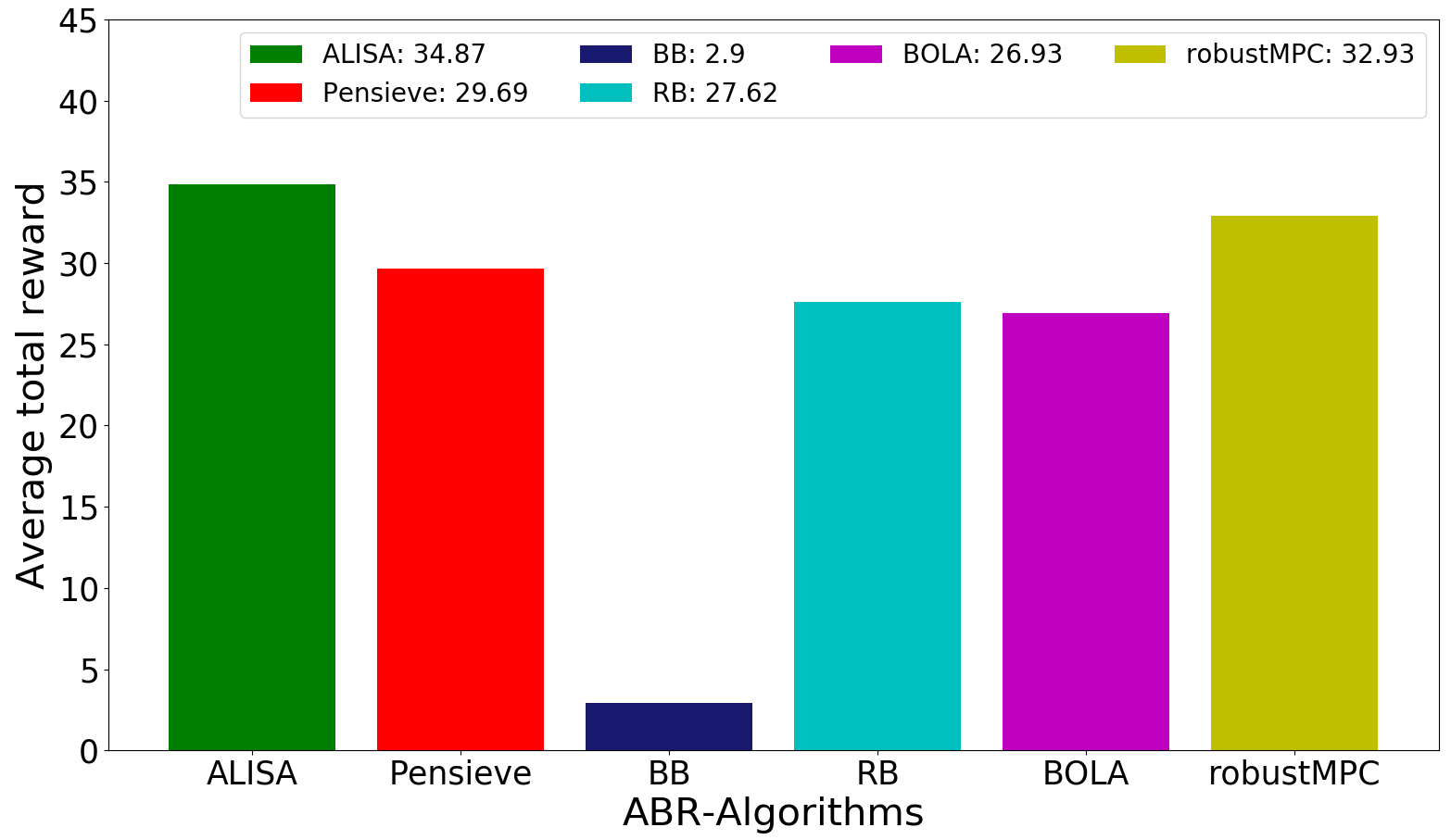}
\caption{Comparison of ALISA over other ABR algorithms with the \(QoE_{linear}\) metric under emulation with 1\% packet loss Equation \ref{eq:qoe}}
\label{fig:bar}
\end{figure}

\begin{figure}
\centering
\includegraphics[width=0.5\textwidth]{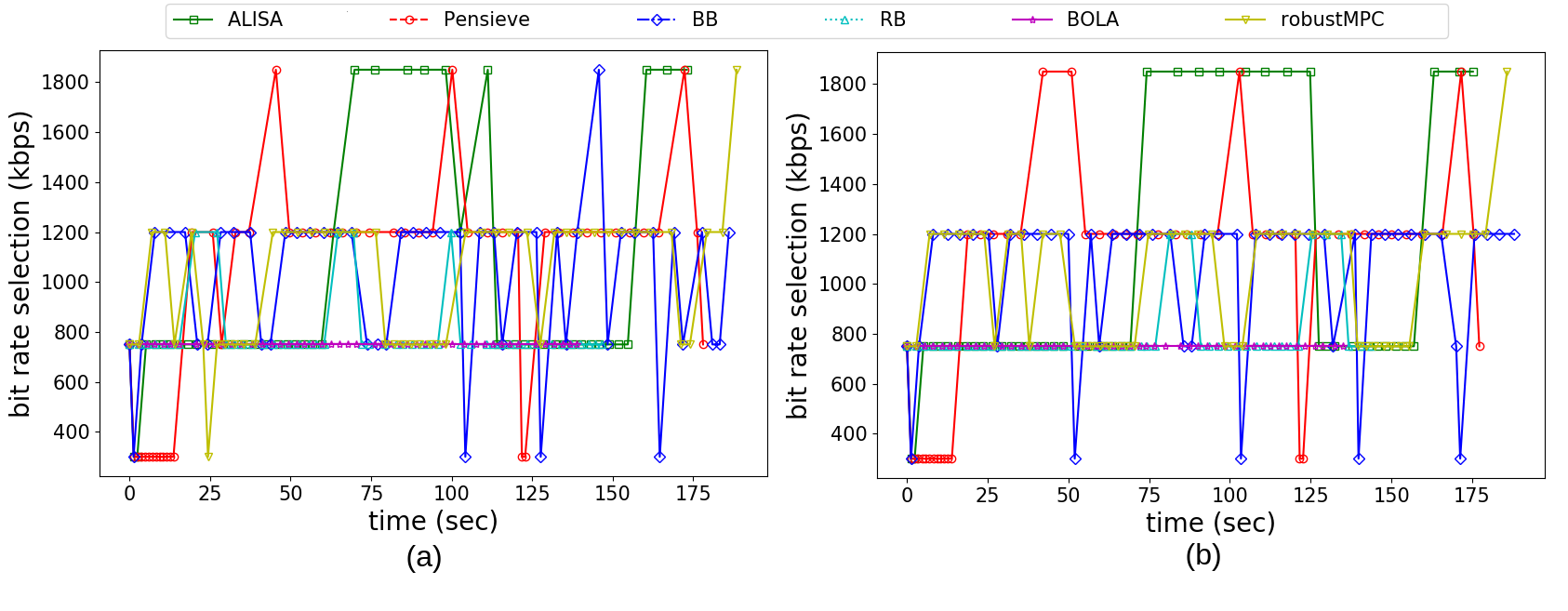}
\caption{Comparison of bit rate selection for ALISA over other ABR algorithms: (a) bitrate selection for sample trace 1, and (b) bitrate selection for sample trace 2 using FCC and Norway trained model for $QoE_{lin}$ metric under emulation with
1\% packet loss (Equation \ref{eq:qoe}).}
\label{fig:bitrate}
\end{figure}

\begin{figure}
\centering
\includegraphics[width=0.5\textwidth]{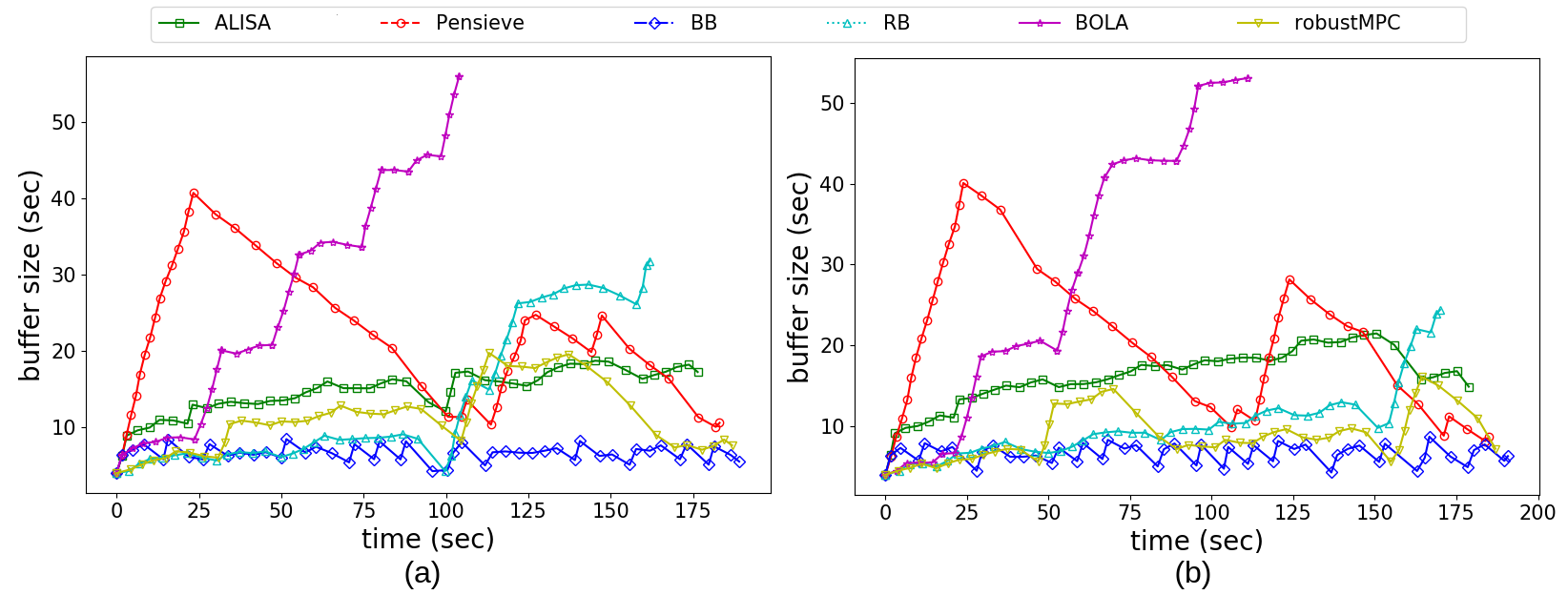}
\caption{Comparison of buffer size selection for ALISA over other ABR algorithms: (a) buffer size selection for sample trace 1, and (b) buffer size selection for sample trace 2 using FCC and Norway trained model for $QoE_{lin}$ metric under emulation with
1\% packet loss (Equation \ref{eq:qoe}).}
\label{fig:buffersize}
\end{figure}

\begin{figure}
\centering
\includegraphics[width=0.5\textwidth]{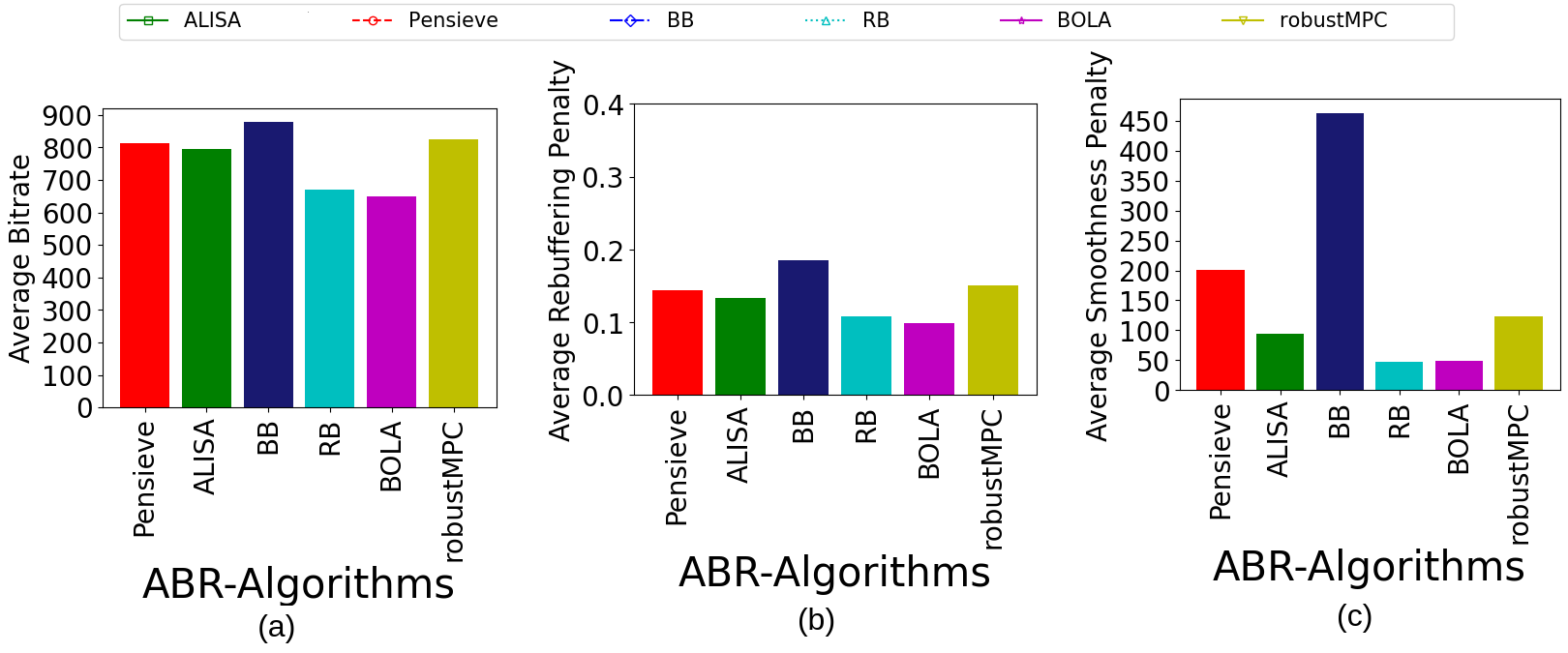}
\caption{Comparing ALISA with existing
ABR algorithms by analyzing their performance on the individual components: (a) average bitrate, (b) average rebuffering penalty, and (c) average smoothness penalty for $QoE_{lin}$ metric under emulation with
1\% packet loss (Equation \ref{eq:qoe}).}
\label{fig:9}
\end{figure}

\subsubsection{Results with Test Data Sets}
We also compare ALISA to several other state-of-the-art ABR algorithms, such as RB, BB, BOLA, and RobustMPC, described in the previous section. We have also compared ALISA with Pensieve, an RL-based basic A3C approach that does not utilize the importance sampling weights. Table \ref{tab:testresultslossless} presents the comparison when there are no losses in the network. Our results show that ALISA achieves a higher QoE on all metrics over all different configurations. ALISA obtains up to 25\% higher QoE than RB, 230\% higher QoE than BB, 30\% higher QoE than BOLA, 25\% higher QoE than RobustMPC and 20\% higher QoE compared to Pensieve  when tested under lossless conditions. This performance translates to lossy conditions as well. We note that ALISA is able to obtain up to 25\%, 28\%, 48\%, and 48\% higher QoE compared to Pensieve  under losses of 0.1\%, 0.5\%, 1\%, and 2\%, respectively. We summarize the remained of our testing QoE metrics for a random packet loss percentage of 0.1\%, 0.5\%, 1\%, and 2\% in Table \ref{tab:testresults.1}, Table \ref{tab:testresults.5}, Table \ref{tab:testresults1} and Table \ref{tab:testresults2}, respectively. These results indicate that ALISA achieves a significantly better performance than many other fixed-rule-based ABR algorithms and also Pensieve. Further, we also visualize the different components of the QoE metric from equation (\ref{eq:qoe}) to understand how ALISA performs better than other ABR algorithms. Figure \ref{fig:rewardcdf} presents the total reward achieved by various
ABR algorithms with $QoE_{lin}$ metric for each trace when
the network is emulated with 0.1\% packet loss. Our results
show that the ALISA algorithm achieves
a higher average QoE of 44.58 as
compared to other ABR algorithms. Figure \ref{fig:bar} presents the
average total reward achieved by various ABR algorithms with
$QoE_{lin}$ metric for each trace when the network is emulated
with 1\%  packet loss. Our results show that the ALISA algorithm achieves a higher average QoE of 34.87 as compared to other ABR algorithms.

Figure \ref{fig:bitrate}(a) and Figure \ref{fig:bitrate}(b)  shows how ALISA can consistently achieve higher bitrates than other methods for random sample traces. This increases the first component of QoE. From Figure \ref{fig:buffersize}(a) and Figure \ref{fig:buffersize}(b), we note that ALISA maintains an adequate buffer size than Pensieve for random sample traces, leading to a decrease in the second component due to moderate bitrates while maintaining the moderate buffer size in equation (\ref{eq:qoe}) hence ALISA provides a lower rebuffer penalty. On the other hand, Pensieve maintains a higher buffer size than ALISA, which leads to high bitrates. The higher bitrates lead to a high rebuffer penalty. Overall, this leads to a higher quality of experience for ALISA over other ABR algorithms.

To understand and demonstrate the better performance of the ALISA, we analyze the individual component of the QoE metric
and present the comparison of various ABR algorithms using
the average playback bitrate, average rebuffering penalty, and
the average smoothness penalty for $QoE_{lin}$ metric under
emulation with $1\%$ packet loss in Figure \ref{fig:9}. Our results show
that most of the ABR algorithms achieve a higher bitrate
except for BOLA and RB in  Figure \ref{fig:9}(a). Due to the selection of a higher bitrate,
several of these algorithms suffer from a rebuffering penalty
where BB and robustMPC have the highest rebuffering penalty in Figure \ref{fig:9}(b).
Similarly, BB also suffers from a high smoothness penalty in  Figure \ref{fig:9}(c). The
ALISA achieves a higher average bit rate
and comparatively smaller rebuffering and smoothness penalty.
The overall impact of these individual components results in
the ALISA achieving an average QoE
higher than the other ABR algorithms. We have observed
similar results for the $QoE_{lin}$, $QoE_{log}$ and   $QoE_{HD}$ metrics under emulation with 0.1\%, 0.5\%, 1\%, and 2\%  packet losses. 

\section{Conclusions}
\label{sec:conclusions}
We have demonstrated how combining importance sampling and structured entropy selection considerably enhances the performance of vanilla A3C approaches (using the Pensieve framework) when used to generate ABR algorithms for video delivery service. By incorporating these methods into our proposed system, ALISA, we are able to consistently achieve up to a 25-48 \% improvement in QoE and even higher in some cases. Additionally, we evaluate our approaches under a broader range of network conditions in terms of packet loss and observe comparable benefits. Finally, we visualize and compare ALISA's bitrate selection and buffer size to those of other ABR algorithms (RB, BB, BOLA, and robustMPC), and shown that ALISA outperforms them in both areas, resulting in an improved QoE. The future work will examine advanced hybrid cloud-edge architectures for ALISA implementation. Additionally, we intend to investigate ALISA in a federated environment in order to take advantage of distributed training across multiple decentralized edge devices.

\section*{Acknowledgment}

This work has been supported by the TCS foundation, India under the TCS research scholar program, 2019-2023.

\bibliographystyle{ieeetr}

\bibliography{cas-refs.bib}

\end{document}